\documentstyle[aps,prb,epsf,preprint]{revtex}
\draft
\begin{document}
\columnsep -.375in
\title{Spatial Correlations in Dynamical Mean Field Theory}

\author{J. Lleweilun Smith and Qimiao Si}
 
\address{Department of Physics, Rice University, Houston,
TX 77251-1892, USA}

\maketitle

\begin{abstract}
We further develop an extended dynamical mean field approach 
introduced earlier. It goes beyond the standard $D=\infty$
dynamical mean field theory by incorporating quantum fluctuations
associated with intersite (RKKY-like) interactions.
This is achieved by scaling the intersite interactions
to the same power in $1/D$ as that for the kinetic terms.
In this approach, a correlated lattice problem is reduced
to a single-impurity Anderson model with additional
self-consistent bosonic baths. 
Here, we formulate the approach in terms of perturbation
expansions. We show that the two-particle vertex functions
are momentum-dependent, while the single-particle
self-energy remains local. In spite of this, the approach is 
conserving. Finally, we also determine the form of a
momentum-dependent dynamical susceptibility; the resulting
expression relates it to the corresponding Weiss field,
local correlation function and (momentum-dependent)
intersite coupling.

\end{abstract}
 
\vskip 0.2 in

\section{\bf Introduction}
\label{sec:intro}

In strongly correlated electron systems, both local and
non-local interactions are important in determining
the nature of the ground state and low lying excitations.
One example illustrating this point comes from Kondo systems,
such as heavy fermions. Here, the competition between the
local Kondo interactions and non-local RKKY interactions
was recognized to play an essential role from 
very early on\cite{Doniach,Jones}. The Kondo effect tends 
to quench local moments altogether, while the RKKY interactions
promote magnetic ordering. What happens when the two processes
are about ``equally'' important is an intriguing question
which remains poorly understood. This question has once
again become centrally important, due to the emergence of a host
of heavy fermion metals lying in the vicinity
of a quantum phase transition\cite{ITP}.

The interplay between local and non-local interactions is also
essential in Mott-Hubbard systems\cite{Edwards}. 
When on-site interactions are strong, their effects can be
thought of as determining the atomic configurations
that lie at low energies. The precise form of the ground state
and low-lying excitations, on the other hand, have to
be determined by the residual intersite couplings between
these low energy configurations.

In general, a separation of electron-electron
interactions into local and non-local ones is not necessarily
well-defined. Local interactions, when combined with
kinetic terms, can lead to effective non-local interactions. 
After all, both RKKY and super-exchange interactions arise
in this fashion. 

In theoretical approaches, however, such a separation 
can often be sharply defined.
In this paper, we are concerned with the dynamical mean
field theory (DMFT)\cite{Georges} which
is formally exact in the limit of infinite dimensions
($D=\infty$)\cite{Metzner}.
The DMFT reduces a correlated lattice problem to a 
self-consistent Anderson impurity model, namely 
a quantum impurity coupled to a self-consistent
fermionic bath.
The interactions between the impurity degrees of freedom
reflect the on-site interactions of the lattice problem;
in this way local quantum fluctuations are retained.
The self-consistent fermionic bath of the
impurity problem reflects the influence, at the
one-particle level, of the rest of the lattice on
the selected (i.e., impurity) site.
All the intersite correlations of the lattice problem,
on the other hand, are neglected. In this sense,
non-local quantum fluctuations are completely lost.

In earlier works, we\cite{Smith1,Smith2} and independently
Kajueter and Kotliar\cite{Kajueter} have extended the
the DMFT such that intersite quantum fluctuations are
treated on an equal footing with local ones.
This extended DMFT reduces a correlated lattice problem
into a novel effective impurity problem, which corresponds
to an Anderson impurity model with additional 
self-consistent bosonic baths. These bosonic baths
reflect the influence,
at the two-particle level, of the rest of the lattice on the
impurity site. Through self-consistency, they keep track of
the intersite quantum fluctuations.
In these earlier works, the mean field equations were derived using
the so-called ``cavity''-method.

The purpose of this paper is to give an alternative formulation
of this extended DMFT using perturbation methods.  We are then 
also able to explicitly
show the conserving nature of this approach. Finally, we also 
derive the expressions for momentum-dependent correlation functions,
which specify how to calculate the correlation functions 
from the corresponding Weiss  fields, local 
correlation functions and (momentum-dependent) intersite interactions.

To be specific, we focus on a one band model,
\begin{eqnarray}
H = && \sum_{i} U n_{i \uparrow} n_{i \downarrow} + 
\sum_{<ij>,\sigma} t_{ij} c_{i\sigma}^{\dagger}c_{j\sigma} \nonumber\\
&&+ {1 \over 2} \sum_{<ij>} v_{ij} (n_{i}-<n>) 
(n_{j}-<n>) + 
{1 \over 2} \sum_{<ij>} J_{ij} \vec{S_{i}} \cdot 
\vec{S_{j}} 
\label{hamiltonian}
\end{eqnarray}
The first two terms alone would correspond to the standard Hubbard model
for a spin $1/2$ band. The third and forth terms are the intersite
density-density ($v_{ij}$) and spin-exchange ($J_{ij}$)
interactions. Here $n_i$ and $\vec{S_i}$ are the density and spin operators
for the $c-$electrons. $<ij>$ labels a pair of nearest-neighbor sites.
For simplicity, we limit both the hopping and intersite-interaction
terms to nearest-neighbor only. 
Generalizing to the case with longer-range hopping and interaction terms 
is straightforward.

In the standard large $D$ approach, the single-electron hopping term is
taken to be of order $1/\sqrt{D}$:
\begin{eqnarray}
t_{<ij>} = t / \sqrt{D}
\label{tij}
\end{eqnarray}
The large $D$ limit leads to an effective single-site problem,
which describes an impurity coupled to a self-consistent
Weiss field. There is one Weiss field for each frequency,
due to the quantum mechanical nature of the hopping term.
The coupling of the impurity to this frequency-dependent
Weiss field can be equivalently described in terms of
a coupling between the impurity and an effective non-interacting
fermionic bath. The bath is fermionic, since the Weiss field 
describes the influence of the rest of sites to the selected impurity
site at the one-particle level.
On the other hand, the intersite interaction terms (both $v_{<ij>}$ and
$J_{<ij>}$) are taken to be of order $1/D$. 
With such scaling, only the static Hartree contributions
survive the large $D$ limit\cite{footnote}. All the quantum fluctuations
associated with these interactions are then neglected.

In the extended DMFT, 
these intersite interaction terms are also scaled to the order 
$1/\sqrt{D}$. The large $D$ limit then leads to a different
effective impurity problem: the impurity is now also coupled 
to the frequency-dependent Weiss field induced by these intersite
interactions. 
The two-particle nature of the intersite 
interactions dictate the bosonic nature of the corresponding
effective baths in the impurity problem. 
As a result,
the effective single-site problem can be 
thought of as an impurity coupled not only to a self-consistent
fermionic bath but also to self-consistent bosonic baths. 

The rest of the paper is organized as follows. In Section 
\ref{mean-field-equations}, we introduce the extended large 
$D$ limit and derive the mean field equations using 
perturbation methods.
We then derive the expressions for the
momentum-dependent correlation functions in Section 
\ref{correlation-function}.
The proof that the approach is conserving is given in
Section \ref{conserving}.
In Section \ref{discussion},
we a) generalize the approach to
multi-band systems as well as to the case of an ordered
state; b) specify an approximate procedure to deal 
with incommensurate spatial fluctuations; and finally
c) compare our approach with others within the general
DMFT framework.

\section{\bf The extended dynamical mean field approach}
\label{mean-field-equations}

In this section, we introduce the extended large $D$ limit and
derive the mean field equations.

The intersite hopping term, $t_{<ij>}$, remains of order 
$1/\sqrt{D}$ as given in Eq. (\ref{tij}). 
The nearest-neighbor interactions are now scaled to the same order,
\begin{eqnarray}
v_{<ij>} = v/\sqrt{D} \nonumber\\
J_{<ij>}=J/\sqrt{D}
\label{Jij}
\end{eqnarray}
In order for the large $D$ limit to be well-defined,
we need to subtract the Hartree contribution
as has already been done in Eq. (\ref{hamiltonian}).

\begin{figure}[h]
\epsfxsize=6.5 in
\centerline{\epsffile{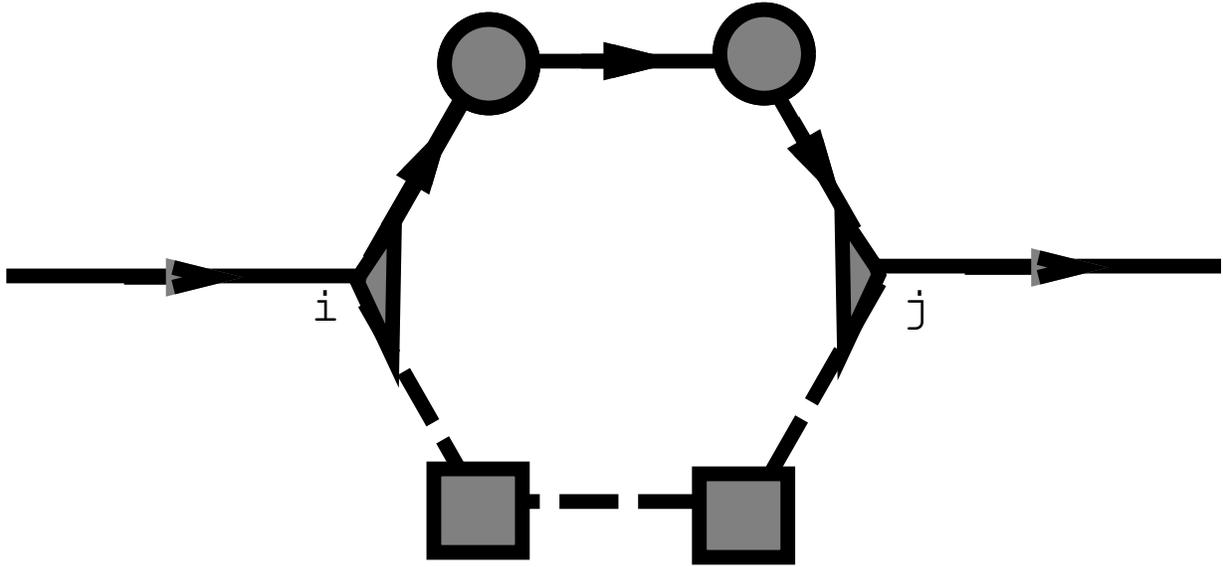}}
\vspace{0.5 in}
\caption{A typical intersite self-energy diagram. A full line represents
a single-particle Green's function, and a dashed line
corresponds to an intersite interaction. This diagram is sub-leading.}  
\label{fig:self_energy_intersite_12-24-98}
\end{figure}

We now establish that the self-energy is still local.
Consider first the self-energy $\Sigma_{<ij>}$.
The Hartree contributions from both $v_{<ij>}$ and $J_{<ij>}$ vanish.
The Fock and high-order contributions to the self-energy
can be written in a skeleton expansion. 
As illustrated in Fig. \ref{fig:self_energy_intersite_12-24-98}, 
any skeleton expansion 
diagram for the self-energy contains at least an intersite interaction
path and a fermion propagator from site $i$ to site $j$.
Both are at least of order $1/\sqrt{D}$. Therefore, $\Sigma_{<ij>} \sim O(1/D)$.
More generally, $\Sigma_{ij} \sim O(1/D)^{||i-j||}$,
where $||i-j||$ is the least number of steps from
site $i$ to site $j$. This implies that the self-energy is 
momentum-independent: $\Sigma ({\bf q}, \omega) =\Sigma_{ii}(\omega)$.

Consider now the on-site self-energy $\Sigma_{ii}(\omega)$.
The only real-space self-energy diagrams that survive the large
$D$ limit have the form illustrated in Fig. 
\ref{fig:self_energy_on_site_12-24-98}a),
as explained in detail in Appendix \ref{sec:self-energy}.
Here a solid line represents the fermion propagator, $G_{ii}$.
A dashed line denotes the intersite interaction (either $v_{ij}$
or $J_{ij}$). The loop formed by two dashed lines enclosing
a solid square represents
either $\chi_{ch,0}^{-1}(\omega)$ or $\chi_{s,0}^{-1}(\omega)$ which,
as explained in Appendix \ref{sec:self-energy},
has the following form,
\begin{eqnarray}
\chi_{ch,0}^{-1} &&= \sum_{ij} v_{i0}v_{0j}(\chi_{ch,ij}
- \chi_{ch,i0}\chi_{ch,0j}/\chi_{ch,00})\nonumber\\
\chi_{s,0}^{-1} &&= \sum_{ij} J_{i0}J_{0j}(\chi_{s,ij} -
\chi_{s,i0}\chi_{s,0j}/\chi_{s,00})
\label{chi_0_selfconsistent}
\end{eqnarray}
where $\chi_{ch}$ and $\chi_{s}$ are the charge and spin susceptibilities
respectively.

The above implies that $\Sigma_{ii}$ can be equivalently calculated
in terms of a local problem with an action of the following form,
\begin{eqnarray}
S^{MF} = &&\int_0^{\beta} d \tau~U
n_{\uparrow}(\tau) n_{\downarrow} (\tau) 
 - \int_{0}^{\beta} d \tau \int_{0}^{\beta} d \tau'
~~[\sum_{\sigma}
c_{\sigma} ^ {\dagger} (\tau) G_0^{-1}(\tau - \tau ')
c_{\sigma}(\tau') + \nonumber\\
&& + :n(\tau): \chi_{ch,0}^{-1}(\tau - \tau'):n(\tau'):
+ \vec{S}(\tau) \cdot \chi_{s,0}^{-1}(\tau - \tau') \vec{S}(\tau') ]
\label{action_selfconsistent}
\end{eqnarray}
where $:n: \equiv n - <n>$. 
The skeleton expansion for the self-energy of this local problem,
$\Sigma_{loc}$, has the form given in Fig. 
\ref{fig:self_energy_on_site_12-24-98}b) 
where a solid line represents $G_{loc}$ and a shaded line
represents either $\chi_{ch,0}^{-1}$ or $\chi_{s,0}^{-1}$.
Since there exists a one-to-one correspondence between the diagrams
in Figs. \ref{fig:self_energy_on_site_12-24-98}a) and 
\ref{fig:self_energy_on_site_12-24-98}b), we have 
\begin{eqnarray}
\Sigma_{loc} = \Sigma_{ii}
\label{sigma-loc-ii}
\end{eqnarray}
if $G_0$ is chosen such that
\begin{eqnarray}
G_{loc} = G_{ii}
\label{G_selfconsistent}
\end{eqnarray}
In standard fashion\cite{Georges}, it can be seen from the Dyson
equations for both the local problem and the lattice problem
that, Eqs. (\ref{sigma-loc-ii},\ref{G_selfconsistent})
are satisfied if the Weiss field is chosen as
\begin{eqnarray}
G_0^{-1}(i \omega_n) &&= i\omega_n + \mu - 
\sum_{ij} t_{i0}t_{0j} [ G_{ij}(i \omega_n) -
G_{i0}(i \omega_n)G_{0j}(i \omega_n)/G_{00}(i \omega_n) ]
\label{G_0_selfconsistent}
\end{eqnarray}

Consider now the higher-order correlation functions.
In a skeleton expansion for any on-site correlation function
of the lattice problem, every diagram also has the
form of Fig. \ref{fig:self_energy_on_site_12-24-98}a) (with even(odd) 
number of fermion loops for any even(odd)-number-particle correlation functions).
Likewise, the skeleton expansion diagrams for the
corresponding local correlation function of the impurity problem
have the structure of Fig. \ref{fig:self_energy_on_site_12-24-98}b). 
As a result Eqs. (\ref{chi_0_selfconsistent},\ref{G_selfconsistent}), 
or equivalently,
Eqs. (\ref{chi_0_selfconsistent},\ref{G_0_selfconsistent}),
also guarantee that all the on-site higher-order correlation functions
of the lattice problem can be calculated from the effective impurity problem.
In particular, $\chi_{ch,loc} = \chi_{ch,ii}$ and $\chi_{s,loc} = \chi_{s,ii}$.

Eqs. (\ref{action_selfconsistent}, \ref{G_0_selfconsistent},
\ref{chi_0_selfconsistent}) form the self-consistency equations. 
The effective impurity problem, defined by the action given in
Eq. (\ref{action_selfconsistent}), can be equivalently written
in terms of an impurity Hamiltonian of the following form,
\begin{eqnarray}
{H}_{imp} = &&
{H_{kin}} +E_d \sum_{\sigma} c_{\sigma}^{\dagger} c_{\sigma}
+  {U } n_{c \uparrow}  n_{c \downarrow} 
+t \sum_{k\sigma} ( c^{\dagger}_{\sigma} \eta_{k\sigma} + H.c. )
\nonumber\\
&&
+F \sum_{q} :n_{c}: (\rho_q + \rho_{-q}^{\dagger}) 
+g \sum_{q} \vec{S}_{c} \cdot
(\vec{\phi}_q + \vec{\phi}_{-q}^{\dagger}) \nonumber\\
H_{kin} = && \sum_{k\sigma} E_k \eta_{k\sigma}^{\dagger}\eta_{k\sigma}
+ \sum_{q} W_q \rho_{q}^{\dagger} \rho_q 
+ \sum_{q} w_q \vec{\phi}_{q}^{\dagger} \cdot \vec{\phi}_q
\label{himp}
\end{eqnarray}
where $E_d=-\mu$ and the parameters $E_k$,$W_q$,$w_q$,$V$,$F$,
and $g$ are given by
\begin{eqnarray}
i\omega_n+\mu -t^{2} \sum_{k} 1/(i\omega_n- E_{k}) &&=
G_0^{-1}(i \omega_n) \nonumber\\
F^{2} \sum_{q} W_{q}/[(i\nu_n)^{2} - W_{q}^{2}] &&=
\chi_{ch,0}^{-1}(i\nu_n)\nonumber\\
g^{2} \sum_{q} w_{q}/[(i\nu_n)^{2} - w_{q}^{2}] &&=
\chi_{s,0}^{-1}(i\nu_n)
\label{parameter_definitions}	
\end{eqnarray}
where $i\omega_n$ and $i\nu_n$ are Matsubara frequencies for fermions
and bosons respectively.

Eq. (\ref{himp}) describes a single-impurity Anderson model
coupled to two additional bosonic bands.
The impurity corresponds to the $c-$orbital.
$\eta_{k\sigma}$ is the usual fermionic bath of the Anderson model,
with a dispersion of $E_k$.
$\rho_q$ is a scalar-bosonic bath. $\vec{\phi}_q$ corresponds to
a vector-bosonic bath. Note that the different components of the
vector boson commute with each other: 
$[\phi_q^{\alpha},\phi_{q'}^{\beta,\dagger}]
= \delta_{\alpha \beta}\delta_{qq'} $ where $\alpha,\beta = x,y,z$.
The dispersions of the bosonic baths are $W_q$ and $w_q$.

As in the usual large $D$ limit, having solved the local problem
we can then calculate the lattice Green's function 
\begin{eqnarray}
G ({\bf k}, \omega) = \frac {1}  {\omega + \mu - \epsilon_k - \Sigma (\omega)}
\label{G_k}
\end{eqnarray}
In the following, we establish that a parallel procedure can be 
carried out for the lattice correlation functions.

\section{\bf Momentum-dependent Correlation Functions}
\label{correlation-function}

We now specify the procedure to calculate the momentum-dependent
correlation functions.
For simplicity, we focus on the spin-spin correlation
function only. The density-density correlation function has
a similar form.

\subsection{Two-particle vertex functions}

\begin{figure}
\centerline{\epsffile{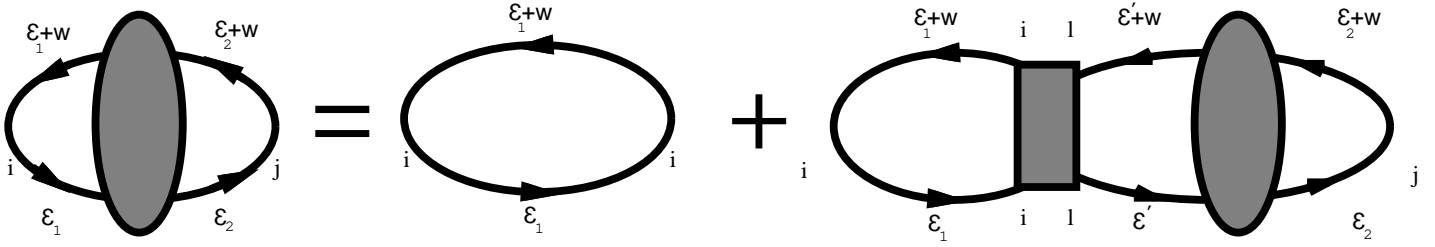}}
\vspace{0.5 in}
\caption{The Bethe-Salpeter equation. Here a bubble with a shaded insertion
represents the susceptibility. The shaded square describes
an irreducible vertex function.
}  
\label{fig:Bethe_Salpeter_12-24-98}
\end{figure}

We first establish the form of two-particle vertex functions.
Consider the spin-spin correlation function,
$\chi_{s} ( {\bf q}, \omega )$, which can be written as
\begin{eqnarray}
\chi_{s} ({\bf q}, \omega) = \int d {\epsilon_1}
d{\epsilon_2} \chi_{ch} (\epsilon_1, \epsilon_2; {\bf q}, \omega )
\label{chi_ch_definition}
\end{eqnarray}
where $\epsilon_1$,$\epsilon_2$ are illustrated in Fig. 
\ref{fig:Bethe_Salpeter_12-24-98}, which also specifies the Bethe-Salpeter 
equation, 
\begin{eqnarray}
\chi_{s,ij} (\epsilon_1, \epsilon_2; \omega)
= \chi^{ph}_{ij} (\epsilon_1; \omega) \delta_{\epsilon_1, \epsilon_2}
+ && \int d \epsilon '  \sum_{l_1,m_1,l_2,m_2}
\chi^{ph}_{il_1im_1}(\epsilon_1; \omega) \times \nonumber\\
&&\times I_{l_1l_2m_1m_2}(\epsilon_1,\epsilon ';\omega)
\chi_{s,l_2jm_2j} (\epsilon ', \epsilon_2; \omega)
\label{chi_ch_vertex_general}
\end{eqnarray}
Here $\chi^{ph}$ represents the particle-hole bubble,
with full fermion propagators.
$I_{l_1m_1l_2m_2}$ is the irreducible vertex function
in the triplet particle-hole channel.
It follows from the counting rules of Appendix \ref{sec:counting}
that only a limited number of contributions on the right hand  
side are of leading order. For the first term on the right
hand side, only the $i=j$ contribution is leading.  
For the second term, only terms with $l_1=m_1=i$ and $l_2=m_2$
contribute. Eq. (\ref{chi_ch_vertex_general}) then leads to
\begin{eqnarray}
\chi_{s}({\bf q},\omega)^{-1}_{\epsilon_1,\epsilon_2}
=\chi^{ph}(\epsilon_1;\omega)^{-1}\delta_{\epsilon_1, \epsilon_2}
- I(\epsilon_1,\epsilon_2 ;{\bf q},\omega)
\label{chi_ch_Iq}  
\end{eqnarray}
where $I(\epsilon_1,\epsilon_2;{\bf q},\omega)
\equiv \sum_j e^{i\vec{q} \cdot \vec{R}_{ij}}
I_{ijij}(\epsilon_1,\epsilon_2;\omega)$. Note that
Eq. (\ref{chi_ch_Iq}) is a matrix equation, 
with $\epsilon_1,\epsilon_2$ specifying matrix elements.

\begin{figure}
\centerline{\epsffile{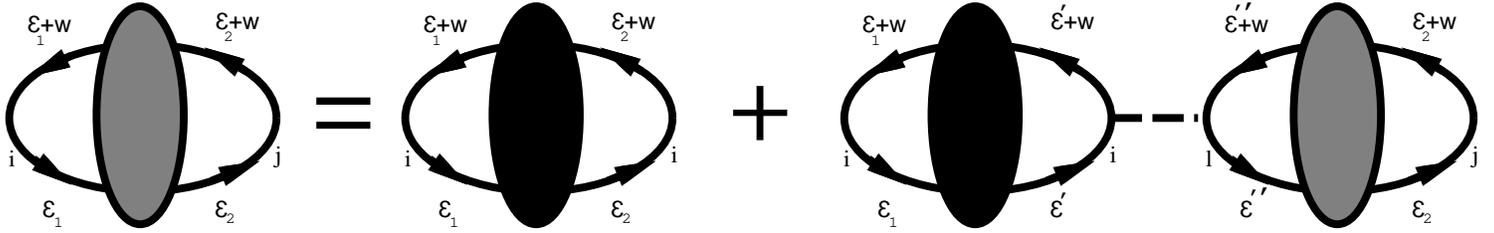}}
\vspace{0.5 in}
\caption{Susceptibility expansion in terms of cumulants.
The bubble with a shaded insertion represents the susceptibility,
while the bubble with a solid insertion corresponds to 
an effective two-particle cumulant.}  
\label{fig:Cumulant_2-24-99}
\end{figure}

We now need to calculate the irreducible vertex function
$I(\epsilon_1,\epsilon_2;{\bf q},\omega)$ in terms of the effective
impurity problem. To do that, we first carry out a
cumulant expansion for the spin-spin correlation function of the
lattice model. A cumulant is introduced in a 
perturbative expansion in terms of $t_{ij}$, $v_{ij}$ and $J_{ij}$. It
represents the corresponding local single-particle or two-particle
Green's function calculated entirely in terms of the on-site (``atomic'')
part of the lattice Hamiltonian.
For our purpose, it is more convenient to introduce an effective spin
cumulant in analogy to the one-particle effective cumulant 
introduced by Metzner\cite{Metzner2}. The effective spin cumulant
is defined as all the diagrams for the on-site spin-spin
correlation function which are irreducible in terms of
cutting any $J_{lm}$ line (where $l$ and $m$ are arbitrary).
Loosely speaking, it is the bare spin cumulant plus all the local decorations.
Denoting the effective spin cumulant as 
$C_{s}(\epsilon_{1}, \epsilon_{2}; \omega)$, the leading order
spin-spin correlation function has the structure illustrated
in Fig. \ref{fig:Cumulant_2-24-99} and can be written as
\begin{eqnarray}
\chi_{s,ij} (\epsilon_1, \epsilon_2;\omega)
= C_{s}(\epsilon_1, \epsilon_2; \omega)
\delta_{ij}
+\sum_l \int d {\epsilon '} \int d \epsilon ''
C_{s}(\epsilon_1, \epsilon ';\omega)
~J_{il} ~\chi_{s,lj} (\epsilon '', \epsilon_2; \omega)
\label{chi_s_chain}
\end{eqnarray}
which is equivalent to
\begin{eqnarray}
\chi_{s}({\bf q},\omega)^{-1}_{\epsilon_1,\epsilon_2}
= C_{s}(\omega)^{-1}_{\epsilon_1,\epsilon_2} - J({\bf q})
\label{chi_ch_vq}
\end{eqnarray}

The leading in $1/D$ contributions to the effective spin cumulant 
contain diagrams of the form illustrated for the on-site self-energy
in Fig. \ref{fig:self_energy_on_site_12-24-98}.
This again follows from the power counting 
rules of Appendix \ref{sec:counting} and can be derived following a
procedure parallel to that of Appendix \ref{sec:self-energy}.
As a result, all the loops involving
intersite interactions are equal to the corresponding Weiss
fields $\chi_{s,0}^{-1}$ and
$\chi_{ch,0}^{-1}$. 

\begin{figure}
\centerline{\epsffile{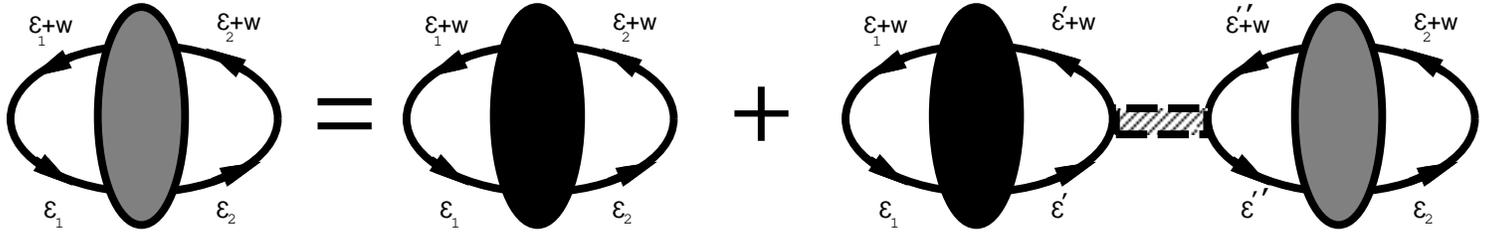}}
\vspace{0.5 in}
\caption{Local susceptibility expansion in terms of cumulants and the
Weiss field.}  
\label{fig:Cumulant_Local}
\end{figure}

The above, in turn implies that the effective spin cumulant 
is equal to the local 
spin-spin correlation functions of the effective impurity problem, Eq. 
(\ref{action_selfconsistent}), which are not reducible in terms of cutting a single 
$\chi_{s,0}^{-1}$ line. We then have the following equation for the local
spin-spin correlation function,
\begin{eqnarray}
\chi_{s,loc}(\epsilon_1, \epsilon_2; \omega) = C_{s}(\epsilon_{1}, 
\epsilon_{2}; \omega) +  \int d \epsilon' \int d \epsilon'' C(\epsilon_{1}, 
\epsilon'; \omega) ~\chi_{s,0}^{-1} (\omega) ~ \chi_s
(\epsilon'', \epsilon_{2}; \omega) 
\label{cumulant-chi-local}
\end{eqnarray}
as illustrated in Fig. \ref{fig:Cumulant_Local}.
Eq. (\ref{cumulant-chi-local}) leads to
\begin{eqnarray}
C_{s}(\omega)^{-1}_{\epsilon_1, \epsilon_2} = 
\chi_{s,loc}(\omega)^{-1}_{\epsilon_1, \epsilon_2} 
+ \chi_{s,0}^{-1}(\omega)
\label{cumulant-chi-local2}
\end{eqnarray}
which specifies how to calculate the effective spin cumulant
from the Weiss field and the local spin-spin correlation function.

Combining Eqs. (\ref{chi_ch_vq},\ref{chi_ch_Iq},\ref{cumulant-chi-local2}),
we derive the following form for the momentum-dependent 
irreducible vertex function,
\begin{eqnarray}
I(\epsilon_1,\epsilon_2;{\bf q},\omega)
&&= \chi^{ph}(\epsilon_1;\omega)^{-1} \delta_{\epsilon_1, \epsilon_2}
- \chi_{s,loc}(\omega)^{-1}_{\epsilon_1, \epsilon_2} - \chi_{s,0}^{-1}(\omega)
 + J({\bf q}) 
\label{Irreducible.vertex}
\end{eqnarray}
as well as the local irreducible vertex function,
\begin{eqnarray}
I_{loc} (\epsilon_1,\epsilon_2 ; \omega)
= \chi^{ph}(\epsilon_1;\omega)^{-1}\delta_{\epsilon_1,\epsilon_2}
- \chi_{s,loc}(\omega)^{-1}_{\epsilon_1, \epsilon_2} - \chi_{s,0}^{-1}(\omega)
\label{I_loc}
\end{eqnarray}

\subsection{\bf Correlation functions}

We now proceed to determine the momentum-dependent correlation
function. Again we focus on the spin-spin correlation function.

Integrating both sides of Eq. (\ref{chi_s_chain})
over both $\epsilon_1$ and $\epsilon_2$ leads to the      
following,
\begin{eqnarray}
\chi_{s,ij} (\omega) = M_{s}(\omega) \delta_{ij}
+\sum_l M_s(\omega) J_{il} \chi_{s,lj} (\omega)
\label{chi_lattice_cumulant}
\end{eqnarray}
where the integrated spin cumulant
$M_s(\omega) \equiv \int d \epsilon_1 \int
d \epsilon_2 C_{s}(\epsilon_1,\epsilon_2;\omega)$.
Eq. (\ref{chi_lattice_cumulant}) yields,
\begin{eqnarray}
\chi_{s} ({\bf q}, \omega) ={M_s(\omega) \over  
{1 - J({\bf q}) M_s(\omega)}}
\label{chi_q_cumulant}
\end{eqnarray}

The integrated spin cumulant $M_{s}(\omega)$ can be determined
from integrating both sides of Eq. (\ref{chi_s_chain})
over both $\epsilon_1$ and $\epsilon_2$,
\begin{eqnarray}
\chi_{s,loc} (\omega) = M_{s}(\omega)
+M_s(\omega) ~ \chi_{s,0}^{-1}(\omega) ~ \chi_{s,loc} (\omega)
\label{chi_loc_cumulant}
\end{eqnarray}
which yields,
\begin{eqnarray}
M_s(\omega) = { \chi_{s,loc} (\omega) \over {1 + \chi_{s,0}^{-1}(\omega) 
\chi_{s,loc}(\omega)}}
\label{eff_cumulant}
\end{eqnarray}

Inserting Eq. (\ref{eff_cumulant}) into Eq. (\ref{chi_q_cumulant})
leads to the final form for the momentum-dependent correlation function,
\begin{eqnarray}
\chi_{s} ({\bf q}, \omega) ={1 \over
{{1/\chi_{s,loc}(\omega)} + \chi_{s,0}^{-1}(\omega) - J({\bf q})}}
\label{chi_q}   
\end{eqnarray}

Eq. (\ref{chi_q}) is one of the most important conclusions
of this paper. It specifies how to calculate the 
momentum-dependent spin susceptibility from
the local spin susceptibility, the Weiss field
($\chi_{s,0}^{-1}$, see Eq. (\ref{chi_0_selfconsistent})),
and the exchange interaction. One check for the
validity of this expression can be seen by re-writing it 
in the following form,
\begin{eqnarray}
\chi_{s} ({\bf q}, \omega) =
\chi_{s,loc} (\omega) + \chi_{s,loc}(\omega) 
[J({\bf q}) - \chi_{s,0}^{-1} (\omega)] \chi_s ({\bf q}, \omega)
\label{chi_q_check}   
\end{eqnarray}
Using the expression for the Weiss field $\chi_{s,0}^{-1}$,
Eq. (\ref{chi_0_selfconsistent}),
one finds that summing both sides of 
Eq. (\ref{chi_q_check}) over ${\bf q}$ leads to 
$\chi_{s,loc} = \chi_{s,loc}$.
An alternative derivation of 
Eq. (\ref{chi_q}) is given in Appendix
\ref{sec:susceptibility}.

\section{\bf Conserving Criteria}
\label{conserving}

We see from the above that the single-particle self-energy is local,
but the two-particle vertex function is momentum-dependent. 
We show in this section that, in spite of this, the approach is
conserving.

\begin{figure}
\centerline{\epsffile{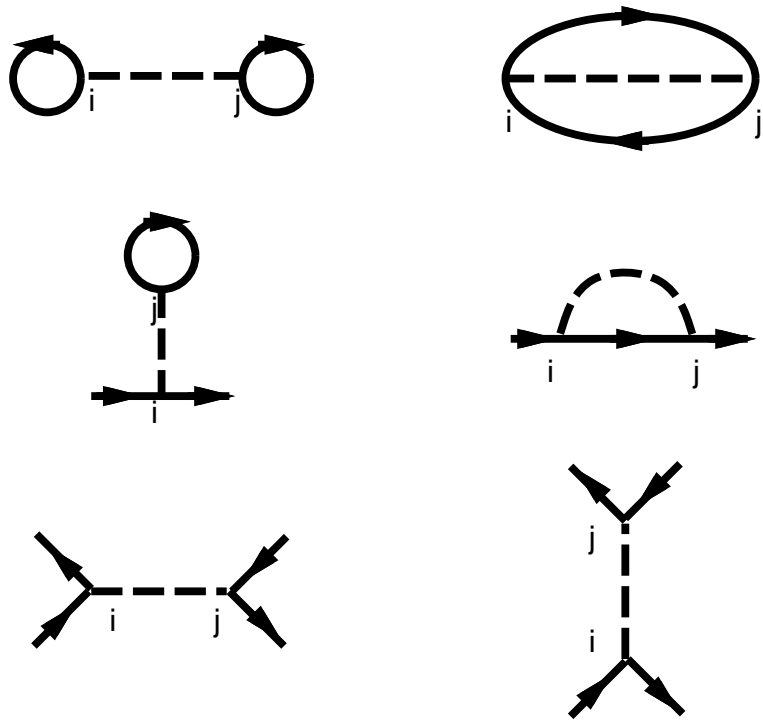}}
\vspace{0.5 in}
\caption{Hartree and Fock contributions to a) the Luttinger-Ward potential;
b) the self-energy; and c) the vertex function.}  
\label{fig:Hartree_Fock_new}
\end{figure}

An approach is conserving 
if both the single-particle self-energy and two-particle
irreducible vertex functions are variational derivatives with
respect to the single-particle Green's functions of the
Luttinger-Ward $\Phi-$potential\cite{Baym}.
The latter is defined as the sum of all closed skeleton diagrams,
and is a functional of the single-particle Green's function
and interaction parameters.
Applying this criterion to our case is somewhat subtle.
The momentum-dependent part of the vertex function
comes from diagrams for the Luttinger-Ward potential
that are sub-leading in $1/D$. These include 
the Hartree and Fock contributions. In fact, due to the usage
of normal-ordered operators 
in the Hamiltonian the Hartree contributions to 
Luttinger-Ward potential and 
to the self-energy, given in Figs. \ref{fig:Hartree_Fock_new}a) and 
\ref{fig:Hartree_Fock_new}b) 
respectively, are identically zero. The corresponding Fock contributions,
also given in Figs. \ref{fig:Hartree_Fock_new}a) and 
\ref{fig:Hartree_Fock_new}b) respectively, are sub-leading in 
$1/D$. At the same time, both the Hartree and Fock contributions
to the two-particle vertex functions,
given in Fig. \ref{fig:Hartree_Fock_new}c), are of leading order.

Because of this subtlety, we first use 
an alternate set of conserving criteria introduced 
in Ref. \cite{Baym-Kadanoff}.
Two conditions are sufficient for an approach to be conserving.
The first states $G_{2}(1,3,1^{+},3^{+}) = G_{2}(3,1,3^{+},1^{+})$,
where $G_2$ is the standard two-particle Green's function.
[In this section we follow the short-hand notation of 
Ref. \cite{Baym-Kadanoff}. For instance, $1$ labels
(${\bf x}_1$,$\tau_1$), and $1^{+}$ labels (${\bf x}_1$,$\tau_1+0^+$).]
This condition is satisfied in our case. 

The second condition relates the single-particle self energy
and the reducible two-particle
vertex function. This condition contains two equations,
one derived from the equation of motion, $\partial_{\tau_{1}}
G(1,1') = [G(1,1'),H]$, and the other its adjoint.
In what follows, we show that the first equation is satisfied;
similar arguments lead to the validity of the second equation.
In addition, for notational simplicity, we consider the case
when only the intersite density-density interaction $v_{ij}$
is finite (the conclusion is unchanged if $J_{ij}$ is also
present). In this case, the condition for the Hamiltonian
given by Eq. (\ref{hamiltonian}) has the following form,
\begin{eqnarray}
\Sigma(1,1') && = -\int d\overline{3} d\overline{5}
d\overline{6} d\overline{8} ~~v_{x_{1},x_{\overline{3}}}~~
[ G(1-\overline{3})~\delta(\overline{3}-1') + G(1,\overline{5})~
G(\overline{3},\overline{6})~
\Gamma(\overline{5},\overline{6},1',\overline{8})
~G(\overline{8},\overline{3}) ]\nonumber\\
&&-\int d \overline{5} d\overline{6} d\overline{8} ~~U~~
[ \delta(1-1')~n_{-\sigma}(1) + G(1,\overline{5})~
G(1,\overline{6}) ~\Gamma(\overline{5},\overline{6},1',\overline{8})
~G(\overline{8},1)]
\label{conserve_equation}
\end{eqnarray}
where
$\Gamma$ is the reducible two-particle vertex function.\cite{footnote2}

Consider first the case when ${\bf x}_{1}$ is nearest neighbor to
${\bf x}_{1'}$. The left-hand side is of order $1/D$ and hence
subleading. The right hand side contains the non-local component
of the vertex function; one might then worry about whether it is
still subleading. That it is so can be seen by 
enumerating the possible spatial locations
of the integration variables and using
the rules of Appendix \ref{sec:counting}.
In fact, the dominant contributions are all of order $1/D$. 
We illustrate this with two examples. Consider 
first the case ${\bf x}_{\overline{3}} = 
{\bf x}_{\overline{6}} ={\bf x}_{\overline{8}} = {\bf x}_{1}$,
and ${\bf x}_{\overline{5}}={\bf x}_{1'}$, which contribute to
the second term on the right hand side of Eq. (\ref{conserve_equation}).
The non-local vertex function $\Gamma (1',1,1',1)$, which is of
order $1/\sqrt{D}$, is accompanied by another non-local Green's
function, $G(1,1')$, which is also of order $1/\sqrt{D}$.
Consider next the contribution from the first term on
the right hand side of Eq. (\ref{conserve_equation}),
with ${\bf x}_{\overline{5}}={\bf x}_{1'}$ and
${\bf x}_{\overline{3}} = 
{\bf x}_{\overline{6}} ={\bf x}_{\overline{8}} = {\bf x}_{l}$,
where ${\bf x}_l$ is nearest-neighbor to ${\bf x}_1$.
Again, the non-local vertex function $\Gamma (1',l,1',l)$,
which is now of order $1/D$, is accompanied by $G(1,1')$
and $v_{x_{1},x_{l}}$, both of which are of order
$1/\sqrt{D}$; summing over $l$ gives rise to a factor $D$,
leading to an overall $1/D$ contribution. We conclude
that to leading order, Eq. (\ref{conserve_equation})
is indeed satisfied. Analogous arguments apply to arbitrary
${\bf x}_{1} \neq {\bf x}_{1'}$.

Consider next ${\bf x}_1={\bf x}_{1'}$.  To the leading order
Eq. (\ref{conserve_equation}) becomes,
\begin{eqnarray}  
\Sigma(1,1)  = && -\int d\overline{3}
~~ v_{x_{1},x_{\overline{3}}}~~
G(1,1) ~G(\overline{3}, \overline{3})
~\Gamma(1,\overline{3},1,\overline{3})
~G(\overline{3}, \overline{3})\nonumber\\
&& -U ~~[n_{-\sigma}(1) + G(1,1)
~G(1,1) 
~\Gamma(1,1,1,1)~G(1,1)]
\label{conserve_equation_loc}
\end{eqnarray}
[For notational simplicity, we have suppressed the time indices.]
Here, the right hand side also contains 
the non-local components of the vertex function
$\Gamma$. We can expand the right hand side in terms of the 
bare intersite interaction by expressing 
$\Gamma$ in terms of the irreducible vertex function $I$
and using Eqs. (\ref{Irreducible.vertex},\ref{I_loc}).
This leads to diagrams which are in one to one correspondence
with those for the on-site self-energy.  Eq. (\ref{conserve_equation_loc}) 
is then satisfied as well.

What we have shown can be recast in terms of
the $\Phi-$derivability of both the self-energy and the
irreducible vertex function. The only unusual feature
is that, we need to keep not only the contributions to
the Luttinger-Ward potential which are leading in $1/D$,
but also the Hartree-Fock-type terms which are formally
subleading. The $D \rightarrow \infty$ limit has to be
taken after variationally differentiating the
Luttinger-Ward potential with respect to the Green's
functions. The process of taking a variational
derivative of a diagram with respect to the
single-particle Green's function can change the order
in $1/D$ of its contribution.

\section{\bf Discussions}
\label{discussion}

\subsection{\bf Multi-band models and ordered states}

The generalization of our approach to multi-band cases is
straightforward. For a two-band extended Hubbard model,
the corresponding mean field equations have already been written
down in Refs. \cite{Smith1,Smith2}.

Another important two-band model is the Kondo lattice model,
\begin{eqnarray}
H =&& \sum_{<ij>,\sigma} t_{ij} c_{i\sigma}^{\dagger}c_{j\sigma}
+ \sum_i J_K \vec{S}_{i} \cdot \vec{s}_{c_ i} 
+\sum_{<ij>} J_{ij} \vec{S}_{i} \cdot \vec{S}_{j} 
\label{kondo-lattice}
\end{eqnarray}
where $\vec{S}_{i}$ denotes an impurity spin at site $i$,
and $\vec{s}_{c_ i}$ represents the spin of conduction ($c-$)
electrons at site $i$. Taking the large D limit, again with 
$t_{ij} = t_0/\sqrt{D}$ and $J_{ij} = J_0/\sqrt{D}$, results
in the following effective impurity action,
\begin{eqnarray}
S^{MF} = && S_{top} 
+\int_0^{\beta} d \tau J_K \vec{S} \cdot \vec{s}_{c} \nonumber\\
&&- \int_{0}^{\beta} d \tau 
\int_{0}^{\beta} d \tau' [ \sum_{\sigma} c_{\sigma} ^ {\dagger} (\tau)
G_0^{-1}(\tau - \tau ') c_{\sigma}(\tau')
+ \vec{S}(\tau) \cdot \chi_{s,0}^{-1}(\tau - \tau') \vec{S}(\tau')]
\label{S-imp-kondo-lattice}
\end{eqnarray}
where $S_{top}$ describes the Berry-phase of the impurity spin.
The Weiss fields $G_0^{-1}$ and $\chi_{s,0}^{-1}$ are determined
by the self-consistency equations as given in Eqs. 
(\ref{G_0_selfconsistent},\ref{chi_0_selfconsistent}).
The effective action can equivalently be written in terms of the
following impurity problem,
\begin{eqnarray}
{H}_{imp} = && 
\sum_{k\sigma} E_k \eta_{k\sigma}^{\dagger}\eta_{k\sigma}
+ \sum_{q} w_q \vec{\phi}_{q}^{\dagger} \cdot \vec{\phi}_q
-\mu \sum_{\sigma} c_{\sigma}^{\dagger}c_{\sigma}
+ t \sum_{k\sigma} ( c^{\dagger}_{\sigma} \eta_{k\sigma} + H.c. ) \nonumber\\
&&+J_K \vec{S} \cdot \vec{s}_{c} 
+ g \sum_{q} \vec{S} \cdot (\vec{\phi}_q + \vec{\phi}_{-q}^{\dagger})
\label{himp-kondo-lattice}
\end{eqnarray}
where $E_{k}$, $t$, $w_{q}$ and $g$ may be determined from 
the Weiss fields $G_0^{-1}$ and $\chi_{s,0}^{-1}$ as specified by
Eq. (\ref{parameter_definitions}).

Finally, we can also extend the approach to a state with long-range
commensurate spatial ordering. This requires taking the normal-ordering,
as specified in Eq. (\ref{hamiltonian}), with a site-dependent average charge or spin
appropriate for the ordered state. The dynamical mean field equations,
Eqs. (\ref{action_selfconsistent},\ref{G_0_selfconsistent},
\ref{chi_0_selfconsistent}) still apply.

\subsection{\bf Incommensurate susceptibilities}

The form of the momentum-dependent correlation function given
by Eq. (\ref{chi_q}) applies to generic ${\bf q}$.
The momentum-dependence is entirely given by that of
the corresponding intersite interaction. It does not
depend on the single-particle dispersion.

The situation is different from the standard large $D$ limit,
where the momentum dependence of the correlation functions
is given entirely by the single-particle dispersion. The latter
is possible for lattices with unbounded bare density of states
-- such as hypercubic lattice -- which contains special
${\bf q}$ such that $\epsilon_q = \sum_{ij} e^{i {\bf q}
\cdot {\bf R}_{ij}} t_{ij}$ is of order $\sqrt{D}$.

Formally, the extended DMFT described here can only be defined
for lattices with a bounded bare density of states (such as Bethe lattice).
It becomes ill-defined for lattices with unbounded density of states.  When 
$\epsilon_q$ is of order $\sqrt{D}$, so is $J({\bf q})$; through the form of
the susceptibility, Eq. (\ref{chi_q}), the system would then
become unstable (when $J({\bf q})$ is positive).

One way to approximately incorporate the incommensurate fluctuations
induced by Fermi-surface features is through the 
(exact) Bethe-Salpeter equation
\begin{eqnarray}  
\chi_{s} ({\bf q}, \omega)^{-1}_{\epsilon_1,\epsilon_2} =
\chi^{ph} (\epsilon_1;{\bf q}, \omega)^{-1}
\delta_{\epsilon_1,\epsilon_2} - I(\epsilon_1,\epsilon_2;
{\bf q},\omega)
\label{chi_approx}
\end{eqnarray}
where $\chi^{ph} (\epsilon_1;{\bf q},\omega)$ 
is the usual particle-hole
susceptibility bubble calculated in terms of the full
single-particle propagators. One uses the
the self-energy and the irreducible vertex function,
Eq. (\ref{Irreducible.vertex}), using the
extended DMFT. At the same time, one uses the 
momentum-dependence of the intersite interactions
in a given system at finite dimensions.
This procedure is of course
no longer systematic.

\subsection{\bf Comparison with other approaches}

A direct $1/D$ expansion has been introduced by Schiller and
Ingersent\cite{Ingersent}. The expansion is carried out for
the Luttinger-Ward $\Phi-$potential\cite{Halvorsen},
i.e., all the closed skeleton diagrams.
The leading $(1/D^0)$ order contributions involve, as usual,
only a single site.
The next-to-leading order $(1/D)$ contributions
come from diagrams involving two sites. A dynamical mean
field description to this level requires
two effective actions describing a single impurity
and a two-impurity cluster, each coupled to its respective
self-consistent medium. Similarly, expanding to order $1/D^n$
requires solving simultaneously effective problems involving one-site,
two-site, and up to $n+1$ site clusters, each embedded
in its own self-consistent medium. 

\begin{figure}
\centerline{\epsffile{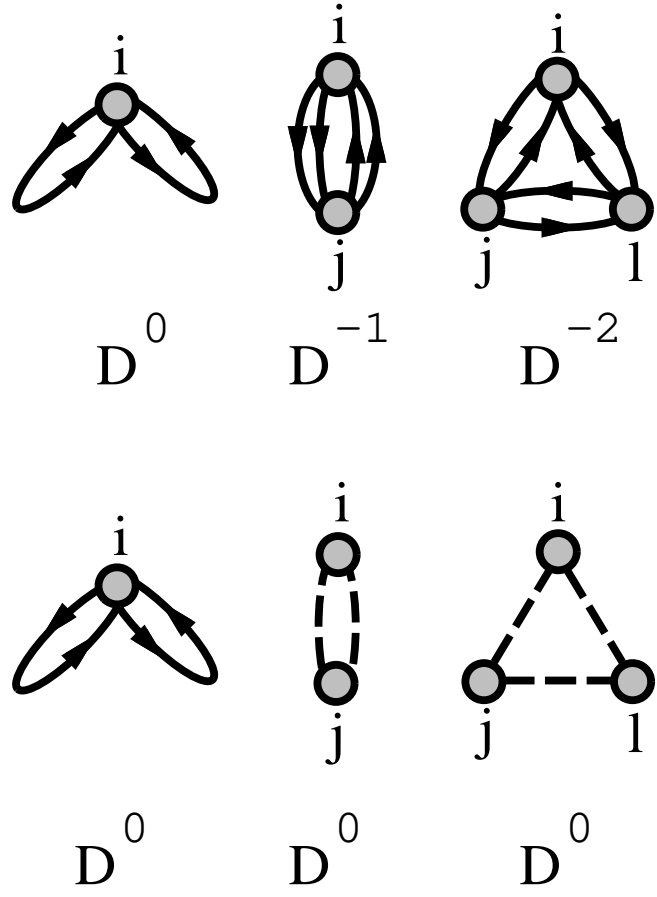}}
\vspace{0.5 in}
\caption{a) single-site, two-site and three-site 
diagrams for the Luttinger-Ward potential in a direct expansion in
$1/D$; b) The corresponding diagrams in the extended DMFT.}  
\label{fig:Comparison_12-24-98v2.ps} 
\end{figure}

The extended DMFT described here 
can be thought of as a conserving
resummation of contributions to all orders in $1/D$. This is
illustrated in Fig. (\ref{fig:Comparison_12-24-98v2.ps}).
A detailed comparison between the two approaches should in principle 
be meaningful when spatial correlations are short-ranged. 

An alternative approach has recently been introduced by Hettler
et al.\cite{Hettler}. In this approach, the Brillouin zone is divided
into $N_c$ regions. For each frequency, one introduces one Weiss field
for each momentum region. The resulting self-consistent problem describes 
an $N_c-$site cluster embedded in these $N_c$ self-consistent Weiss
fields. For $N_c=1$, it reduces to the standard large $D$ DMFT.
The precise relationship between their approach and the extended DMFT
described here is unclear.

Finally, our approach also has some similarities with the dynamical
mean field theory of random spin systems. 
Bray and Moore\cite{BrayMoore} considered the quantum Sherrington-Kirkpatrick
model, in which the exchange coupling is infinite-ranged and has a Gaussian
distribution of mean zero and variance scaled to $J^2/N$
where $N$ is the size of the system. Carrying out disorder averaging
using replicas, as usual, leads to a problem with four-spin interactions;
each spin now carries a replica index.
Since the exchange couplings associated with different bonds are
uncorrelated, the four spin interaction has the form of two spins
at different times $\tau$ and $\tau'$, from any site,
interact with two other spins, also at $\tau$ and $\tau'$,
at every other site. Taking the $N \rightarrow \infty$ limit
then leads to a single-site problem with retarded spin-spin
interactions. In the paramagnetic phase, the effective impurity
problem has a similar form as the 
spin part of Eq. (\ref{action_selfconsistent}), with a self-consistency
equation which is also similar -- though not identical -- to
Eq. (\ref{chi_0_selfconsistent}). Generalizing\cite{Subir,Anirvan}
to the case when conduction electrons are also present 
gives rise to mean field equations similar to those described here.
The form of correlation functions is of course very different.

\section{\bf Conclusions}

To summarize, we have given a perturbative derivation of an
extended dynamical mean field theory. This approach goes
beyond the standard $D=\infty$ DMFT by incorporating the
quantum fluctuations associated with intersite interactions.

The self-consistent impurity problem has the form of an
Anderson impurity model with additional bosonic baths.
This is a novel kind of impurity problem, and is of interest
in its own right. The role of a scalar boson bath is to introduce
additional screening, thereby enhancing orthogonality.
The precise consequences of the enhanced screening depends on
the form of the spectral function of the bosons.
Ref. \cite{Smith1} analyzed an impurity 
problem containing both a scalar boson bath with a spectral function
of $\omega^A$ and a conduction electron band with a regular
density of states at its Fermi energy.
The correlation functions have mean-field exponents.
The effect of an anisotropic vector-boson bath is similar to
the scalar case. The effect of an isotropic 
vector-boson bath is more complex. 
We and independently Sengupta
have carried out\cite{Smith2,Anirvan2} a renormalization group
analysis of spins coupled both to a regular 
conduction electron band ($J_K$) and to a vector-boson bath
($g$) with a spectral function of $\omega^{A}$. In the 
sub-ohmic case ($A<1$), there exists a critical point separating
a phase characterized by the fixed point at $J_K^*=\infty,g^*=0$
and another by $J_K^*=0, g^* \sim \sqrt{1-A}$.
The critical exponents are anomalous\cite{Anirvan2}.
In the absence of conduction electrons, the impurity problem
corresponds to a spin coupled to vector bosons alone.
Sachdev and Ye\cite{Sachdev-Ye} solved such a model in the
large N limit (see also Ref. \cite{Parcollet}). Finally, we note
that quantum Monte Carlo methods have recently been developed
for this type of impurity problems\cite{Marcelo}.

The momentum-dependent correlation functions in the extended DMFT
have the general form given by Eq. (\ref{chi_q}). While different
in details, it is similar to the phenomenological expression
recently introduced in Refs. \cite{Aeppli,Coleman} in the context
of the dynamical spin susceptibility in a heavy fermion metal
($CeCu_{6-x}Au_x$) close to a zero-temperature
phase transition. The experiments\cite{Aeppli,Stockert}
at the critical concentration are suggestive of a dynamical spin
susceptibility $[J({\bf q})+f(\omega,T)]^{-1}$ over the entire
Brillouin zone, where the frequency and temperature dependence
in $f(\omega,T)$ show an anomalous exponent. Whether a quantum
critical point with an anomalous
exponent in the dynamical susceptibility can emerge in the extended
DMFT described here is an exciting open question.

Finally, the extended DMFT described here should be applicable 
to quantitatively study the short, but finite, ranged
dynamical fluctuations in paramagnetic phases
(whether or not the ground state is ordered). Consider,
for example, the spin fluctuations in the paramagnetic phase
of a Mott insulator. In the standard large $D$ limit, these
fluctuations reflect simply isolated local moments:
They contain no damping and are featureless in
momentum space. In the extended DMFT, the self-consistent
problem introduces mode coupling and hence damping. In addition,
through Eq. (\ref{chi_q})
the dynamical spin susceptibility is expected to be peaked
at antiferromagnetic wavevectors. This approach is 
particularly useful at temperatures where the correlation length
is short; here approaches taking into account only long-wave-length
fluctuations would break down. Quantitative calculations 
using the extended DMFT
will perhaps allow a detailed understanding of the neutron
scattering results in both undoped and doped Mott insulators,
such as $V_2O_3$\cite{Bao} where the exchange interactions
are not particularly large.

\acknowledgments

We would like to thank G. Kotliar and D. Vollhardt for useful discussions.
This work has been supported
in part by NSF Grant No. DMR-9712626, Research Corporation,
and A. P. Sloan Foundation.

\newpage

\appendix

\section{\bf Power Counting Rules}
\label{sec:counting} 

The order in $1/D$ of the correlation functions and vertex functions
can be determined by analyzing the diagrams in real space.

Consider first the single particle Green's function $G_{ij}$
and vertex functions $I_{ijij}$ and $\Gamma_{ijij}$. In any 
real space Feynman diagrams for these quantities,
it takes at least $||i-j||$ number of fermion propagation or
intersite interaction steps. Therefore, $G_{ij} \sim I_{ijij}
\sim \Gamma_{ijij} \sim O({1 \over D})^{||i-j||/2}$.

Consider next the vertex functions involving three 
independent sites, 
$I(i,j,i,l)$ and $\Gamma(i,j,i,l)$. Every diagram involves at
least three independent paths of fermion propagators and/or
interaction lines. Therefore, $I(i,j,i,l) \sim \Gamma(i,j,i,l) \sim 
o(1/D)^{M_3/2}$ where 
$M_3=\min \left (
||i-j||+||i-l||,||i-j||+||j-l||,||i-l||+||j-l|| \right )$.

Finally, consider the vertex functions involving four independent
sites, $I(i,j,l,m)$ and $\Gamma(i,j,l,m)$. Every diagram involves
at least four independent paths of fermion propagators and/or
interaction lines. As a result,
$I(i,j,l,m) \sim \Gamma(i,j,l,m) \sim o({1 \over D})^{M_4/2}$ 
where $M_4 = \min ( X_1, X_2, X_3, X_4 )$. Here,
$X_1=||i-j||+||i-m||+||l-m||$, $X_2=||j-i||+||j-l||+||j-m||$,
$X_3=||l-i||+||l-j||+||l-m||$ and $X_4=||m-i||+||m-j||+||m-l||$.

\section{\bf On-site Self-energy}
\label{sec:self-energy} 

\begin{figure}[h]
\centerline{\epsffile{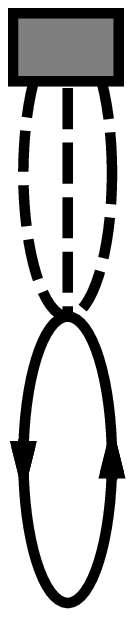}}
\vspace{0.5 in}
\caption{A diagram for the on-site self-energy. It connects the selected
site to the rest of the lattice by three interaction lines, making it
subleading.}
\label{fig:self_energy_forbidden}
\end{figure}

Consider an arbitrary skeleton expansion diagram for the 
on-site self-energy for site $0$, $\Sigma_{00}(\omega)$.
From the counting rules of Appendix 
\ref{sec:counting}, only local fermion propagators appear.
The only non-local terms involve intersite interactions.
These intersite interaction terms can be grouped into separate
loops, each starting at site $0$ and returning to site $0$ as is
illustrated in Fig. \ref{fig:self_energy_on_site_12-24-98}a). Note that, no 
loop can return to site $0$ more than once. For instance, a contribution
given in Fig. \ref{fig:self_energy_forbidden} is subleading.

The analytic expression for each loop can be determined
as follows. The beginning and ending interaction
lines give a product $J_{0l}J_{m0}$, where $l$ and $m$
are arbitrary sites nearest-neighbor to site 0.
The solid square
then represents a correlation function involving $\vec{S}_l$ and
$\vec{S}_m$. Given that site $0$ is excluded from anywhere in
the solid square, this correlation function has to be evaluated
in terms of $H^{(0)}$, defined as the original
Hamiltonian Eq. (\ref{hamiltonian}) with site $0$ excluded.
As a result, this loop can be written as
\begin{eqnarray}  
\sum_{lm}J_{0l}J_{m0} \chi_{lm}^{(0)}
\label{interaction-loop}
\end{eqnarray}
where $\chi_{lm}^{(0)} = <T_{\tau}\vec{S}_l(\tau)\cdot
\vec{S}_m(\tau')>_{H^{(0)}}$, which can be determined as follows.
For $\chi_{lm} = <T_{\tau}\vec{S}_l(\tau)\cdot \vec{S}_m(\tau')>_{H}$,
the cumulant expansion given in Eq. (\ref{chi_s_chain}) implies
$\chi_{lm} = \chi_{ll}\chi_{lm}'\chi_{mm}$, where 
$\chi_{lm}' \equiv \sum_{paths} J_{ll_1}\chi_{l_1l_1}
J_{l_1l_2}\chi_{l_2l_2} ...\chi_{l_n l_n} J_{l_n m} $
and $[l,l_1,l_2, ...,l_n,m]$ labels a non-self-retracing 
path from site $l$ to site $m$. This in turn implies
$\chi_{lm}^{(0)} = \chi_{lm} - \chi_{ll} \chi_{l0}'
\chi_{00}\chi_{0m}'\chi_{mm}$. Therefore,
\begin{eqnarray}  
\chi_{lm}^{(0)} = \chi_{lm} - \chi_{l0}\chi_{0m}/\chi_{00}
\label{chi^0}
\end{eqnarray}
leading to the expression for the spin Weiss field,
$\chi_{s,0}^{-1}$, given in Eq. (\ref{chi_0_selfconsistent}).

Similarly, an interaction chain generated by $v_{ij}$
corresponds to the charge Weiss field,
$\chi_{ch,0}^{-1}$, given in Eq. (\ref{chi_0_selfconsistent}).

\section{\bf Alternative derivation of the momentum-dependent
susceptibility}
\label{sec:susceptibility}

In this section we present an alternative derivation for the momentum-dependent
susceptibility $\chi({\bf q},\omega)$, Eq. (\ref{chi_q}). 

We re-write Eq. (\ref{chi_0_selfconsistent}) in momentum-space,
\begin{eqnarray}
\chi_{0}(\omega)^{-1} = \sum_{q} J^2 ({\bf q}) \chi ({\bf q},\omega)
- [ \sum_{q} J({\bf q}) \chi ({\bf q},\omega ) ]^2 / \chi_{loc}(\omega)
\label{Eq:Chi_ch}
\end{eqnarray}
In addition, we use Eq. (\ref{chi_loc_cumulant}), i.e.,
\begin{eqnarray}
\chi({\bf q},\omega) = \frac{1}{M(\omega)^{-1} - J({\bf q})}
\label{Eq:Chi_cumulant}
\end{eqnarray}

We now substitute Eq. (\ref{Eq:Chi_cumulant}) into Eq. (\ref{Eq:Chi_ch}).
By using
\begin{equation}
\sum_{q} \frac{J({\bf q})}{M(\omega)^{-1} - J({\bf q})}
= -1 + \chi_{loc}(\omega)
\label{Eq:v(k)}
\end{equation}
and
\begin{eqnarray}
\sum_{q} \frac{J({\bf q})^2}{M(\omega )^{-1} - J({\bf q})} 
= M(\omega)^{-1} ( -1 + \chi_{loc}(\omega))
\label{Eq:v(k)^2}
\end{eqnarray}
we obtain $\chi_0(\omega)^{-1} = M(\omega)^{-1} - \chi_{loc}(\omega)^{-1}$,
i.e.,
\begin{eqnarray}
M(\omega)^{-1} = \chi_{loc}(\omega)^{-1} + \chi_0(\omega)^{-1}
\label{Eq:M_Chi}
\end{eqnarray}
Inserting Eq. (\ref{Eq:M_Chi}) into Eq. (\ref{Eq:Chi_cumulant}) then
leads to Eq. (\ref{chi_q}).

Note that,  Eq. (\ref{chi_q}) reduces to the correct result for generic 
${\bf q}$ in the standard large $D$ limit
where $J({\bf q}) = 0$ and $\chi_0^{-1} = 0$.


\vskip 1.0in

\begin{figure}
\epsfxsize=5.0 in
\centerline{\epsffile{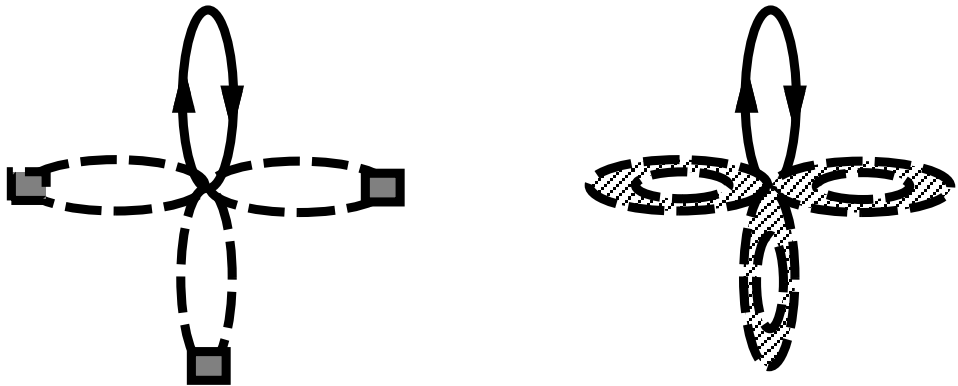}}
\vspace{0.5 in}
\caption{ a) An on-site self-energy diagram of the lattice problem.
b) The corresponding local self-energy diagram of the effective
impurity problem. A shaded line corresponds to a Weiss field.}
\label{fig:self_energy_on_site_12-24-98}
\end{figure}

\newpage

\vskip 1.5 in

\newpage

\vskip 1.0in

\newpage

\vskip 1.5 in

\begin{references}
\bibitem{footnote2} We note that the Fock self-energy diagram
(for ${\bf x}_{1}$ nn to ${\bf x}_{1'}$) 
$v_{x_{1},x_{1'}}G(1,1')$ appears on the right hand side and is 
subleading ($\propto 1/D$).
\end{references}
\end{document}